\documentclass[nofootinbib,floatfix,prd,preprint]{revtex4} %eqsecnum,
\usepackage{graphicx}
\newcommand{\sfig}[2]{
\includegraphics[width=#2]{#1}
		}
\newcommand{\Sfig}[2]{
	\begin{figure}[thbp]
	\sfig{#1.eps}{0.9\columnwidth}
	\caption{{#2}}
	\label{fig:#1}
	\end{figure}
}
\newcommand{\DoubleFig}[3]{
	\begin{figure}[thbp]
\centerline{\hbox{ \hspace{0.0in} 
    \includegraphics[width=0.45\columnwidth]{#1.eps}
   \hspace{0.2in}
    \includegraphics[width=0.45\columnwidth]{#2.eps}
     }
  }
 		 \vspace{9pt}
		\caption{\small #3}
		\label{fig:#1}
        \end{figure}
}

\newcommand{\rf}[1]{\ref{fig:#1}}
\def\mtwo{m_{200}}

\def\be{\begin{equation}}
\def\ee{\end{equation}}
\def\bea{\begin{eqnarray}}
\def\eea{\end{eqnarray}}
\newcommand{\vs}{\nonumber\\*}
\newcommand{\ec}[1]{Eq.~(\ref{eq:#1})}

\newcommand{\eql}[1]{\label{eq:#1}}

\def\fun#1#2{\lower3.6pt\vbox{\baselineskip0pt\lineskip.9pt
  \ialign{$\mathsurround=0pt#1\hfil##\hfil$\crcr#2\crcr\sim\crcr}}}

\def\ga{\mathrel{\mathpalette\fun >}}
\begin{document}
%\baselineskip=24pt
%\twocolumn[\hsize\textwidth\columnwidth\hsize\csname @twocolumnfalse\endcsname
%\pagestyle{empty}
%\begin{center}
%\rightline{{\large DRAFT} (Ewan; July 28, 1999)}
%\bigskip
%\rightline{FERMILAB--Pub--}
%\rightline{astro-ph/0002360}
%\rightline{submitted to {\it Phys. Rev. Lett.}}

%\vspace{.2in}
\title{CMB-Cluster Lensing} 

%\vspace{.2in}
\author{Scott Dodelson$^{1,2}$}
%\vspace{.2in}

\affiliation{NASA/Fermilab Astrophysics Center
Fermi National Accelerator Laboratory, Batavia, IL~~60510-0500}
\affiliation{Department of Astronomy \& Astrophysics, The University of Chicago, 
Chicago, IL~~60637-1433}
\email{dodelson@fnal.gov}

\date{\today}
%\smallskip
\begin{abstract}
Clusters of galaxies are powerful cosmological probes, particularly if
their masses can be determined. One possibility for mass determination is
to study the cosmic microwave background (CMB) on small angular scales
and observe deviations from a pure gradient due to lensing of massive clusters. 
I show that, neglecting contamination,
this technique has the power to determine cluster masses very accurately, in
agreement with estimates by Seljak and Zaldarriaga (1999). However, the intrinsic small
scale structure of the CMB significantly degrades this power. The resulting mass constraints
are useless unless one imposes a prior on the concentration parameter $c$. With even a modest 
prior on $c$, an ambitious CMB experiment
($0.5'$ resolution and $1\mu K$ per pixel) could determine masses of high redshift ($z>0.5$) clusters
with $\sim 30\%
$ accuracy.
\end{abstract}
\maketitle

\section{Introduction} 

Clusters of galaxies are powerful cosmological probes. The
abundance of clusters is very sensitive to the amplitude of fluctuations in the
mass density field. For example, the
present-day mass distribution of clusters offers perhaps the 
cleanest way to measure the
normalization of the matter power spectrum~\cite{Eke1996,Henry1997,
ViaLid1999,ReiBoh2002,Ikebe2002,Schuecker:2002ti,Pierpaoli2003,Allen2003,Viana2003,Dodelson,Rozo2004}. As we gain the ability to probe the
cluster abundance at high redshifts, we can hope to measure the evolution of
this normalization, an evolution which is sensitive to the dark energy and its
equation of state~\cite{HMH2001,Wang:1998gt,Schuecker:2002yj}.

Because the mass function varies so rapidly with mass, accurate mass
determination is critical if cluster constraints are to realize their potential.
Finding clusters is not enough: even if a method finds all clusters and returns
no false positives, it still does not necessarily have constraining power.
It is quite possible that one technique will be optimal for finding clusters
while another is a more powerful probe of cluster masses. For example, radio surveys
which can measure the Sunyaev-Zel'dovich effect are very promising ways of
detecting clusters even at high redshift~\cite{White:2003af}. But they do not necessarily yield
an accurate mass estimator. In contrast, weak gravitational lensing has
drawbacks as a cluster finder~\cite{MetzWhi2001,Whitevan2002} but 
is potentially a powerful way of
measuring mass~\cite{KaiSqu1993,SquKai1996,Hoekstra2001,Hoekstra2003,
Schneider2003,Dod2003}.

While lensing of background galaxies has been studied in depth over the
last decade, lensing of the cosmic microwave background (CMB) has not been
explored as carefully, mainly because the angular resolution needed to see the
predicted signal is only now being attained or 
planned~\cite{ACT,ALMA,Stark1998}. Here I discuss how
accurately cluster masses can be determined with observations of
small angle ($\sim 1'$) observations of the CMB.

On these very small scales, to a first approximation the background primordial 
CMB is a gradient. 
The photons passing close to the center of the cluster 
get deflected so that they appear to be originating further away from the
cluster center. On the cool side of the gradient, this means the temperature is
slightly hotter than it would appear without lensing, and on the hot side
slightly cooler. So, subtracting off the dipole, the lensing pattern is quite
distinctive~\cite{Seljak:1999zn,Zal2000}, as shown in Fig.~\rf{pattern}.

\Sfig{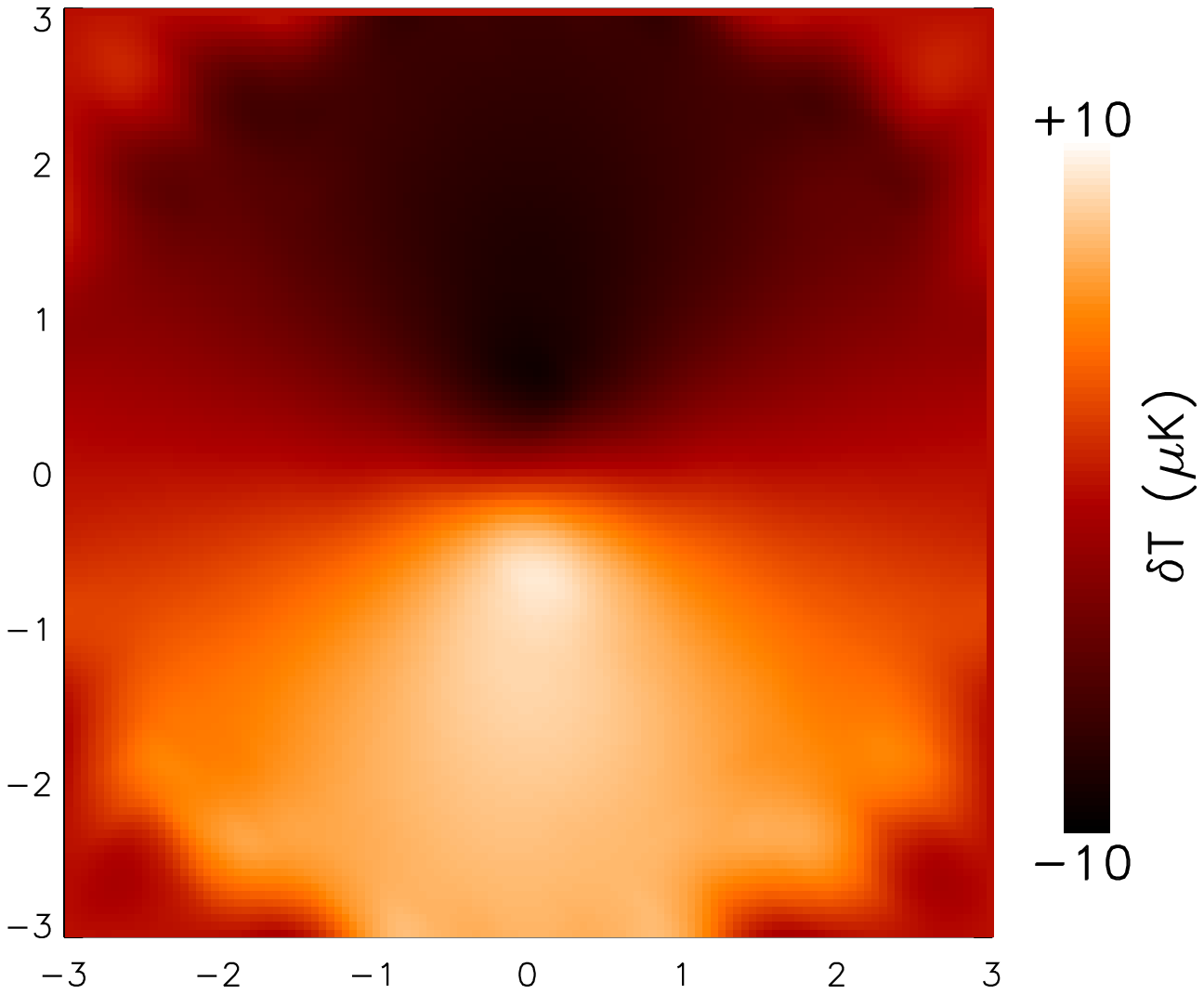}{Idealized temperature pattern due to lensing of a cluster. Here, the background
CMB is assumed to be a pure gradient with strength $T_y=13\mu$K arcmin$^{-1}$ and this gradient has been removed.
The cluster, at redshift 1 with a mass
$\mtwo=10^{15} h^{-1} M_\odot$, has an NFW profile. The area shown is within the virial radius of $3.3'$.
Jaggedness at the edge of the map is an indication of finite size $(0.5')$ of
the (noiseless) pixels.}

Although there are many possible contaminants to this signal, its distinctive
morphology allows one to hope that the signal can be extracted from any sources
of noise~\cite{Seljak:1999zn,HolKos,Vale2004}. Here I examine how accurately CMB lensing can determine the masses of
clusters, accounting for the small scale structure present in the CMB, i.e., the fact that
the background CMB is {\it not} a pure gradient. There are two 
sources of this small scale structure: one is the power which remains in the damping tail
of the primordial ($z\sim1100$) spectrum and the other is lensing by structures along the
line of sight. Figure~\rf{cldt} gives an indication of the strength of the signal and the
contributions from the CMB, which I will call {\it CMB noise}. The CMB noise dwarfs the signal
on ``large'' scales. We will see that this significantly impairs the constraining power of CMB-cluster lensing.

\Sfig{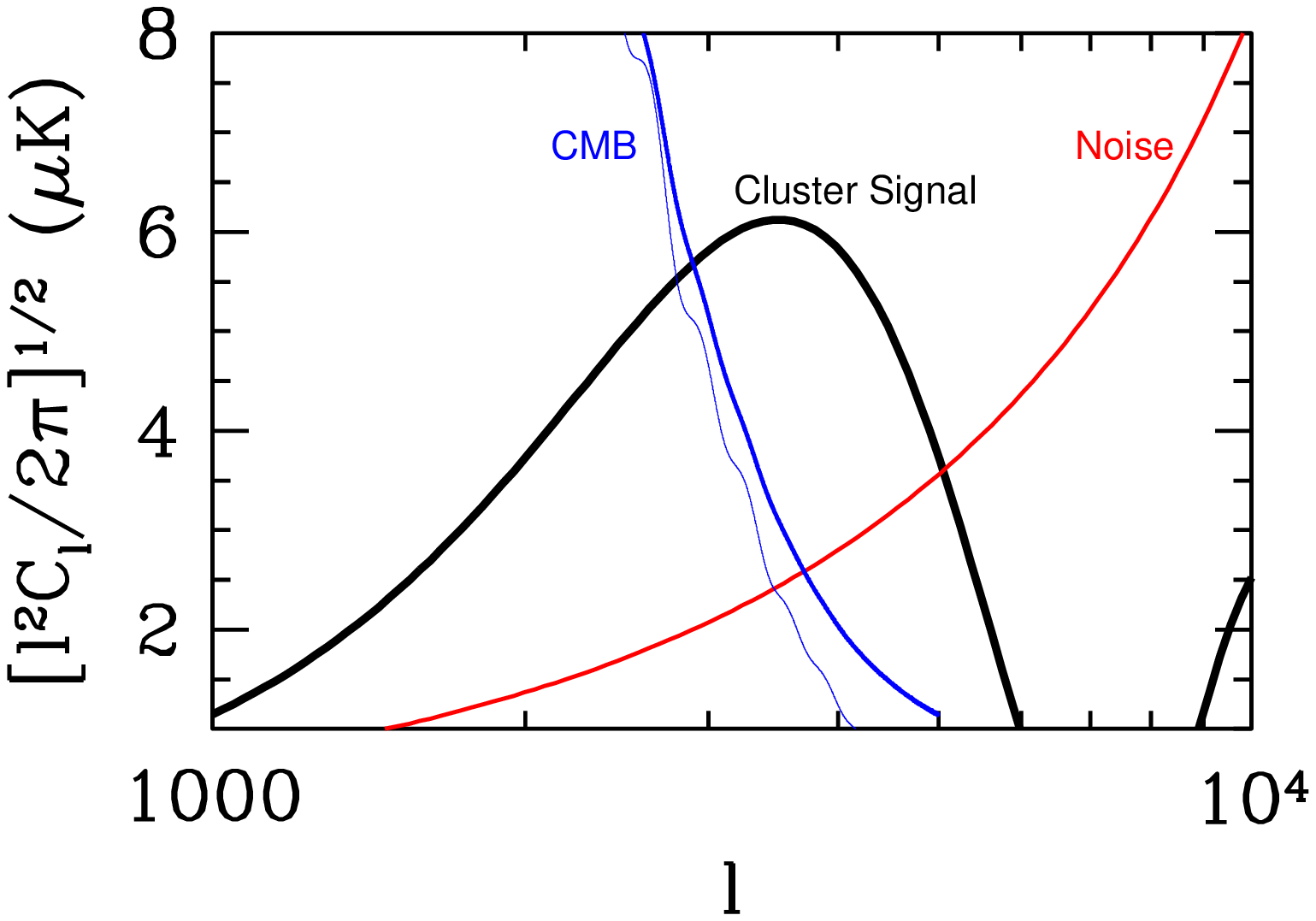}{Anticipated signal from the lensing of a massive cluster and noise in $l$-space. The cluster has 
mass $10^{15} h^{-1} M_\odot$
and is at redshift $1$. The light CMB curve is without lensing along the line of sight; the heavy CMB line
includes this effect. The instrumental noise corresponds to a $0.5'$ resolution experiment
with $1\mu$K noise per pixel.}

Section II describes the model for the mass
density I will use throughout and the ensuing constraints if CMB noise
were absent. Section III shows how CMB noise pollutes the signal and which
modes retain their power. 
Figures~\rf{cont_cmb1d.5}-\rf{dncontcijp_m10z1} summarize the main results of the paper and 
can be understood without
reading the text. Finally, in the conclusion, I emphasize some of the simplifications 
made throughout.

\section{Constraints Neglecting CMB Noise}

Numerical simulations~\cite{Navarro:1997he,Navarro:1996iw} find that clusters obey a Navarro-Frenk-White (NFW)
profile,
\be
\rho(r) = {A\over r (cr+r_v)^2}
\ee
with two parameters the virial radius $r_v$ and concentration $c$. 
The virial radius
is the radius within which the enclosed mass, $\mtwo$, is $200$ times the average mass
in a critical density universe in the same volume. Therefore, it is convenient to use $\mtwo$
instead of the virial radius as the second parameter,
\be
\mtwo =  {800\pi\over 3} \rho_{\rm cr}(z) r_v^3
\ee
where the evolving critical density $\rho_{\rm cr}(z)$ is expressed in terms of the Hubble rate, $3H^2(z)/8\pi G$.
The amplitude of the density is then
\be
A = {\mtwo c^2\over 4 [ \ln(1+c) - c/(1+c) ] }
.\ee
Throughout I will take $c=5$ since this is a typical value emerging from 
simulations~\cite{Bullock2001}, although it does not reflect the changes of $c$ with
redshift or the (slight) changes with mass.

Given this mass distribution, we can ask how the observed signal $\tilde T$ in the CMB 
is affected by the lens. Lensing
can be included by Taylor expanding the temperature field around the unlensed field $T$:
\bea
\tilde T(\vec\theta) &=& T(\vec\theta-\vec{\delta\theta}) \simeq T(\vec\theta) - \vec{\delta\theta}
\cdot {\partial T \over \partial\vec\theta}\vs
&=& -\vec{\delta\theta}_{\rm cl} \cdot {\partial T \over \partial\vec\theta} 
+ \left(  T(\vec\theta) - \vec{\delta\theta}_{\rm lss}
\cdot {\partial T \over \partial\vec\theta}\right)
\eea
where $\vec{\delta\theta}$ is the deflection angle due to lensing along the line of sight
due to the cluster ($_{\rm cl}$ subscript) and unassociated large scale structure ($_{\rm lss}$ subscript). 
Seljak and Zaldarriaga~\cite{Seljak:1999zn,Zal2000} realized that on small scales the primoridial CMB has little
structure so $\partial T/\partial\vec\theta$ in the first term can be set to a constant. 
Aligning the $y$-axis with this vector and calling the constant $T_y$ leads to
\be
\tilde T(\vec\theta) \simeq - {\delta\theta}_{\rm cl,y}(\vec\theta) T_y
+ \left(  T(\vec\theta) - \vec{\delta\theta}_{\rm lss}
\cdot {\partial T \over \partial\vec\theta}\right) .
\eql{dipapp}\ee
To be clear, $\delta\theta_{\rm cl,y}(\vec\theta)$ is the
$y$-component of the deflection angle due to the cluster at angular position $\vec\theta$, and 
$T_y$ is the primordial gradient at the center of the cluster, the derivative of the temperature field with respect
to $\theta_y$. The term in parentheses in \ec{dipapp} is the lensed CMB temperature with mean zero
and variance determined by the set of $C_l$'s appropriate for the chosen cosmology.
Note from Fig.~\rf{cldt} that lensing due to large scale structure gives additional power to
the CMB at the scales of interest.

The deflection angle due to the NFW profile is known to be~\cite{Brainerd,DodSta}
\be
\vec\delta\theta_{\rm cl} (\vec\theta) = -{16\pi G A\over c r_v} {\vec\theta\over\theta} {d_{SL} \over d_S}
g(d_L\theta c/r_v)
\ee
where the distances are angular diameter distances (e.g. 
the angular diameter distance between the source and the lens $d_{SL}$
is
$(\chi_S-\chi_L)/(1+z_s)$ in a flat universe where $\chi$ is the comoving
distance out to the relevant redshift). The function $g$ is of order unity
and peaks when its argument is equal to $1.3$:
\be
g(x) = {1\over x} \cases{ \ln(x/2) + {\ln( x/[ 1-\sqrt{1-x^2}] \over
	\sqrt{1-x^2} }) & $x<1$\cr
	\ln(x/2) + {\pi/2 - {\rm arcsin}(1/x) \over
	\sqrt{x^2-1} } & $x>1$\cr
	}.
\ee

Figure~\rf{pattern} shows the temperature pattern on the sky -- with the 
background gradient
removed -- due to lensing of an NFW cluster with mass $10^{15} h^{-1} M_\odot$ at
redshift 1. Note the distinctive double lobe pattern
on either side of the cluster. Here I have set the background gradient $(T_y)$
to $\sigma_{\rm grad} = 13\mu$K/arcmin, its rms value in a flat cosmology with
$\Omega_\Lambda=0.7, \Omega_b=0.04, h=0.7, \sigma_8=0.9, \tau=0.1$, and primordial 
spectral index $n=1$. The hot and cold spots on either side of the cluster
are then $9\mu$K an arcmin away from the cluster center, dropping off to $7\mu$K
at the virial radius of $3.3'$.
 
\Sfig{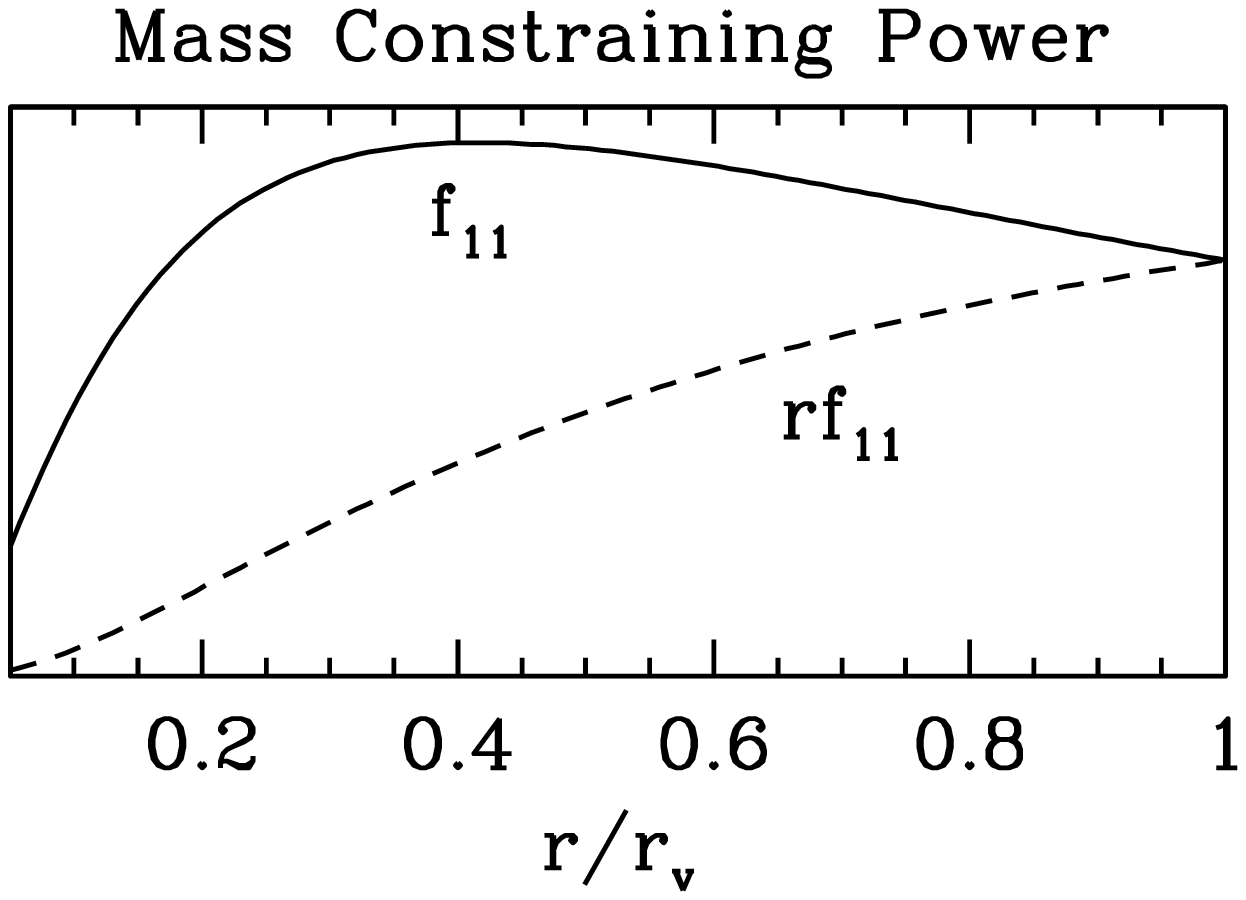}{Contributions to the mass constraint as a function of distance from the cluster
center. Although the largest individual contribution comes from pixels much closer to the center than
the virial radius as indicated by the solid curve,
the pixels at the edge contribute the most in total because there are more of them. This is in the unrealistic
case where only uniform instrumental noise is considered.}

The mild drop of the signal means that the outermost regions of the cluster contribute most
to the mass constraints when only instrumental noise is considered. To see this and to contrast with the
more realistic situation when other sources of noise are included later on, consider the Fisher matrix
which determines the constraints on the parameters $\mtwo$ and $c$:
\be
F_{\alpha\beta} = \sigma_{\rm grad}^2 \sum_{i,j=1}^N {\partial \delta\theta_{\rm cl,y}(\theta_i) \over \partial \lambda_\alpha} 
\left(C^{-1}\right)_{ij} 
{\partial \delta\theta_{\rm cl,y}(\theta_j) \over \partial \lambda_\beta}
.\ee
Here $\alpha,\beta$ label the two parameters ($\lambda_1=\ln\mtwo,\lambda_2=\ln(c)$), 
while $i,j$, which are summed over, label the $N$ pixels
in the map. If the noise matrix is diagonal with uniform noise $\sigma_n$ in all
pixels, which might be true if only instrumental noise is
considered, then the fractional constraint on the mass for example reduces to a single sum over all pixels:
\be
F_{11} = {\sigma_{\rm grad}^2 \over \sigma_n^2}
\sum_{i=1}^N \left({\partial \delta\theta_{\rm cl,y}(\theta_i)  \over \partial \ln\mtwo}\right)^2 \equiv \sum_{i=1}^N f_{11,i}
.\eql{fisher}\ee
That is, each pixel contributes an amount $f_{11,i}$ to the mass constraint. Figure~\rf{littlef} shows
that although the largest contribution comes from the pixels a distance $0.4r_v$ from the cluster center,
when we sum over all pixels in an annulus [$\propto \int dr r f_{11}(r)$], the annulus with the largest radius
has the most constraining power.

\Sfig{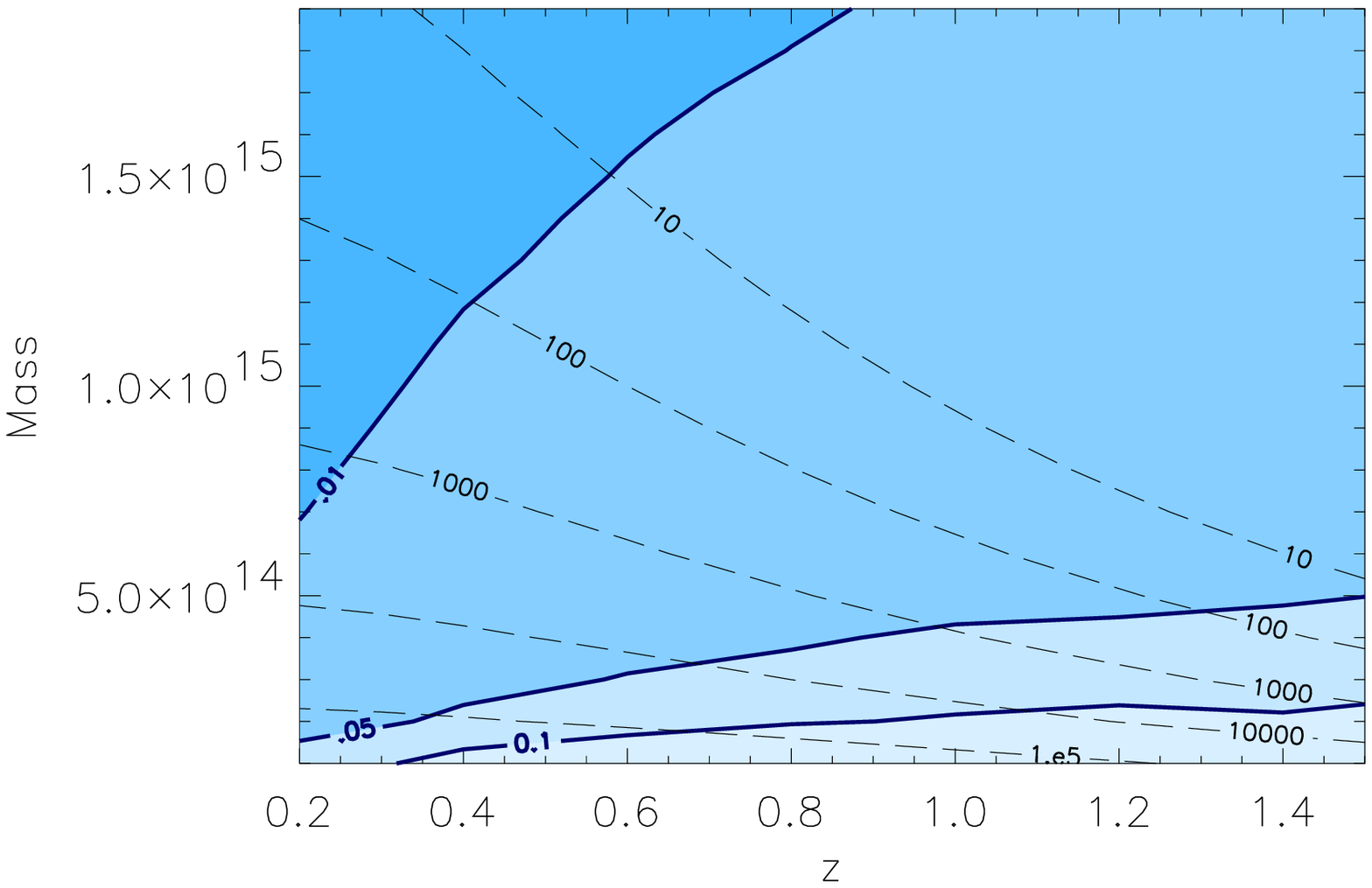}{Shaded contours show $1$-$\sigma$ projections of $\delta M/M$ of a cluster from a CMB
experiment with noise per pixel of 1$\mu$K and pixel size $0.5'$. These
constraints do {\it not} include CMB noise. Dashed lines indicate values for
the cumulative number of clusters (total number above the given mass and
redshift) expected in the standard $\Lambda$CDM
cosmology ($\sigma_8=0.85$).}

If this conclusion held up as more sources of noise were included, it would have dramatic implications
for which clusters were best suited to be studied by CMB lensing. To see this, I will examine the constraints
on the mass and concentration of a cluster by using all pixels out to the virial radius. Beyond the virial
radius, the profile begins to deviate from an NFW profile. While we have no way of knowing in advance
where the virial radius is, one can imagine fitting for all pixels within say $10'$ and getting non-negligible
signal only from distances smaller than the virial radius. As we will see in \S III, this discussion is
academic because CMB noise renders pixels near the virial radius much less useful.

With this strategy, we would constrain the mass of a cluster at lower redshift more tightly than
one (of the same mass) at high redshift because the low redshift object 
subtends a larger fraction of the sky. We can see this in Figure~\rf{cont_nolss} which shows the $1$-$\sigma$
constraints on $\mtwo$ from an idealized CMB experiment with $\sigma_n=1\mu$K and pixel size $\Delta=0.5'$.
The constraints in this unrealistic situation are astounding, percent level errors
on the mass, even after marginalizing over the concentration. Note the trend that
for fixed mass, the constraints do indeed get stronger as the cluster moves to
lower redshift. 
The dashed lines in the figure show the cumulative number of
clusters expected at each mass and redshift. For example, the point
$z=0.2,\mtwo=5\times10^{14} h^{-1} M_\odot$ is intersected (roughly) by
the contour labeled ``10000.'' In this cosmology then, we expect $10^4$ clusters
over the whole sky with mass greater than $5\times10^{14} h^{-1} M_\odot$ at
redshifts greater than $0.2$. (These contours were obtained using the Jenkins mass
function~\cite{Jenkins2001}.)

\section{Constraints Including CMB Noise}

To include noise from the CMB, I generate a set of $C_l$'s for the chosen cosmology including lensing
due to large scale structure using CMBFAST~\cite{cmbfast}. The new covariance matrix is then
\be
C = C^{\rm noise} + \sum_l {2l+1\over 4\pi} C_l P_l(\cos\theta_{ij})
\ee
where $\theta_{ij}$ is the angular distance between pixels $i$ and $j$, and $C^{\rm noise}$
includes instrumental noise, assumed to be diagonal with variance $\sigma_n^2$.

In the presence of this new source of noise, it is useful again to ask which modes
contribute most to the mass constraint. Now though, since the covariance matrix is not
diagonal, we need to generalize \ec{fisher}.
If we diagonalize the full covariance matrix
\be
W = U^t C^{-1} U,
\ee
with the eigenvalues contained in the diagonal matrix $W$
then the Fisher matrix of \ec{fisher} can still be written as a sum of weights, but now the sum is not
over pixels but over modes. That is, the contribution of a given mode to the mass constraint is
\be
f_{11,i} = W_i \sigma_{\rm grad}^2 \left[ \sum_j {\partial \delta\theta_{\rm cl,y}(\theta_j)\over \partial \ln\mtwo} U_{ji}  \right]^2
.\ee
The eigenvector associated with this mode is $\sum_j U_{ji} T(\theta_j)$.

\Sfig{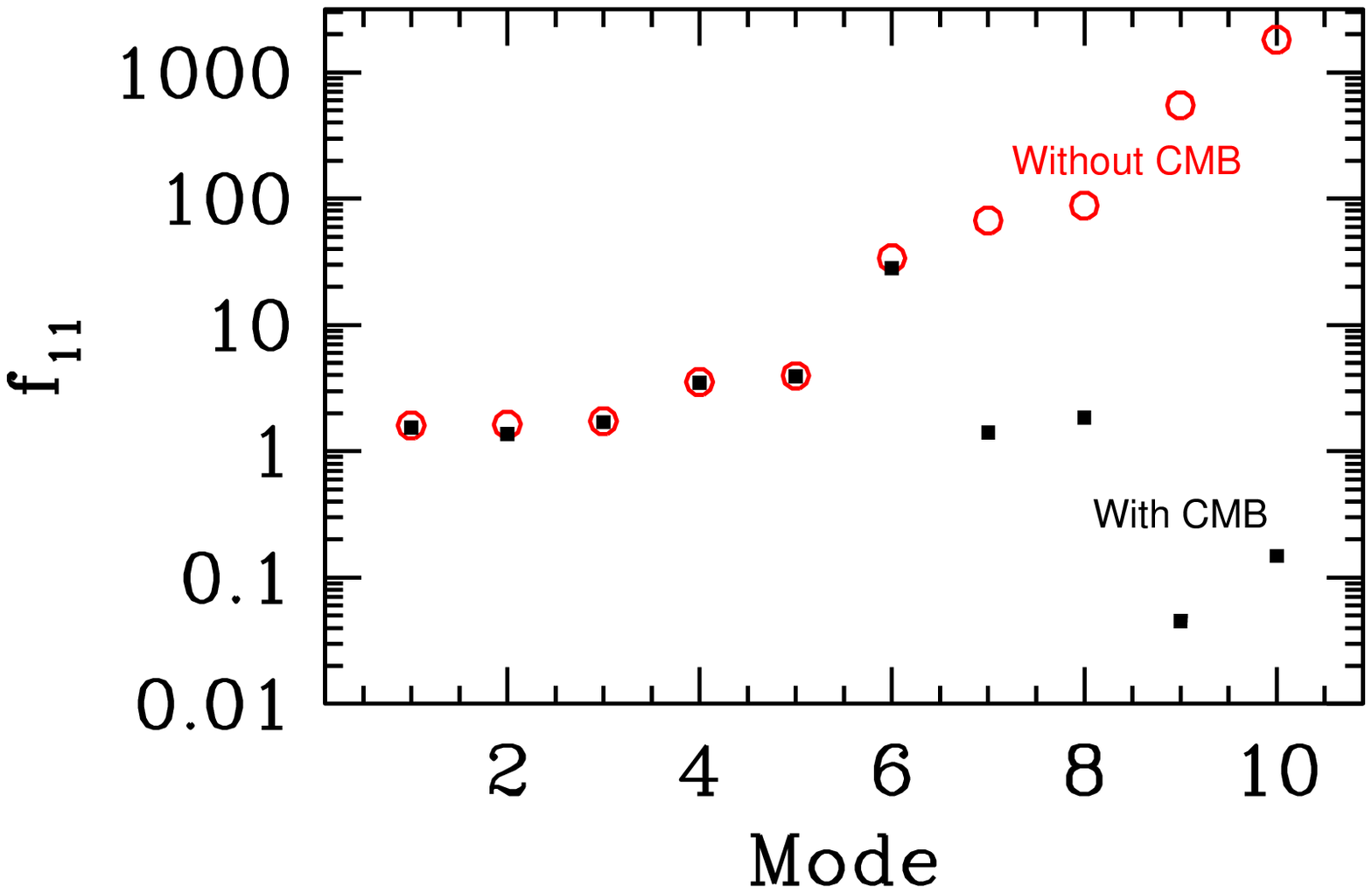}{Constraining power of each of the top $10$ modes when CMB noise is 
not included (red open circles) and when it is (black solid squares). The modes are linear
combinations of the pixels. While of order ten modes are useful in the absence of CMB noise,
only one mode (mode 6 here) contributes significantly to the mass constraints when CMB
noise is taken into account.}

Fig.~\rf{comtmp} illustrates how the signal in many modes
is contaminated by CMB noise. Consider first the (red) circles which show $f_{11}$
for all the modes in the {\it absence} of CMB noise. The mode with the highest constraining power
is depicted in the left panel of Fig.~\rf{u17}: it is essentially a gradient.  
Indeed, you can quickly estimate the mass constraint (with fixed concentration) from this mode: 
$f_{11}$ is of order $2000$, leading to $\delta \mtwo/\mtwo \sim 1/\sqrt{2000} \sim $ a few percent, in
agreement with the result for a cluster with $\mtwo=10^{15} h^{-1} M_\odot$ at $z=1$ shown in Fig.~\rf{cont_nolss}.
The filled squares in Fig.~\rf{comtmp} portray a much different picture when CMB noise is accounted for. 
Now there is only one mode with $f_{11}>10$; it is shown in the right panel of Fig.~\rf{u17}.
It is sensitive to the difference in the cluster signal in the outer and inner parts 
of the cluster. This
mode cannot be reproduced by the CMB because the CMB lacks the necessary small scale power, 
but the signal
in this mode is also quite a bit smaller. Recall from Fig.~\rf{pattern} that the 
signal changes only mildly
from its peak to its value at the virial radius.

\DoubleFig{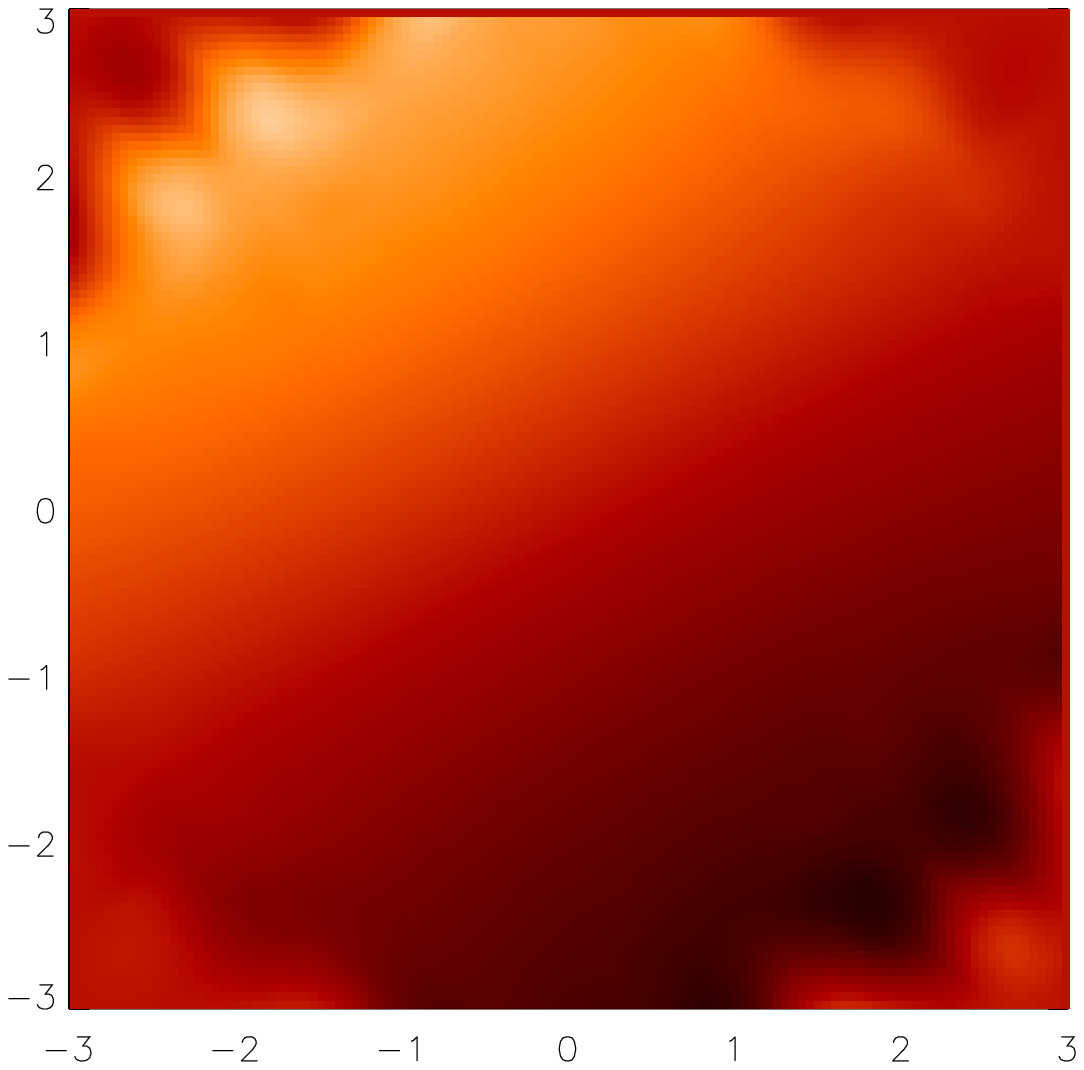}{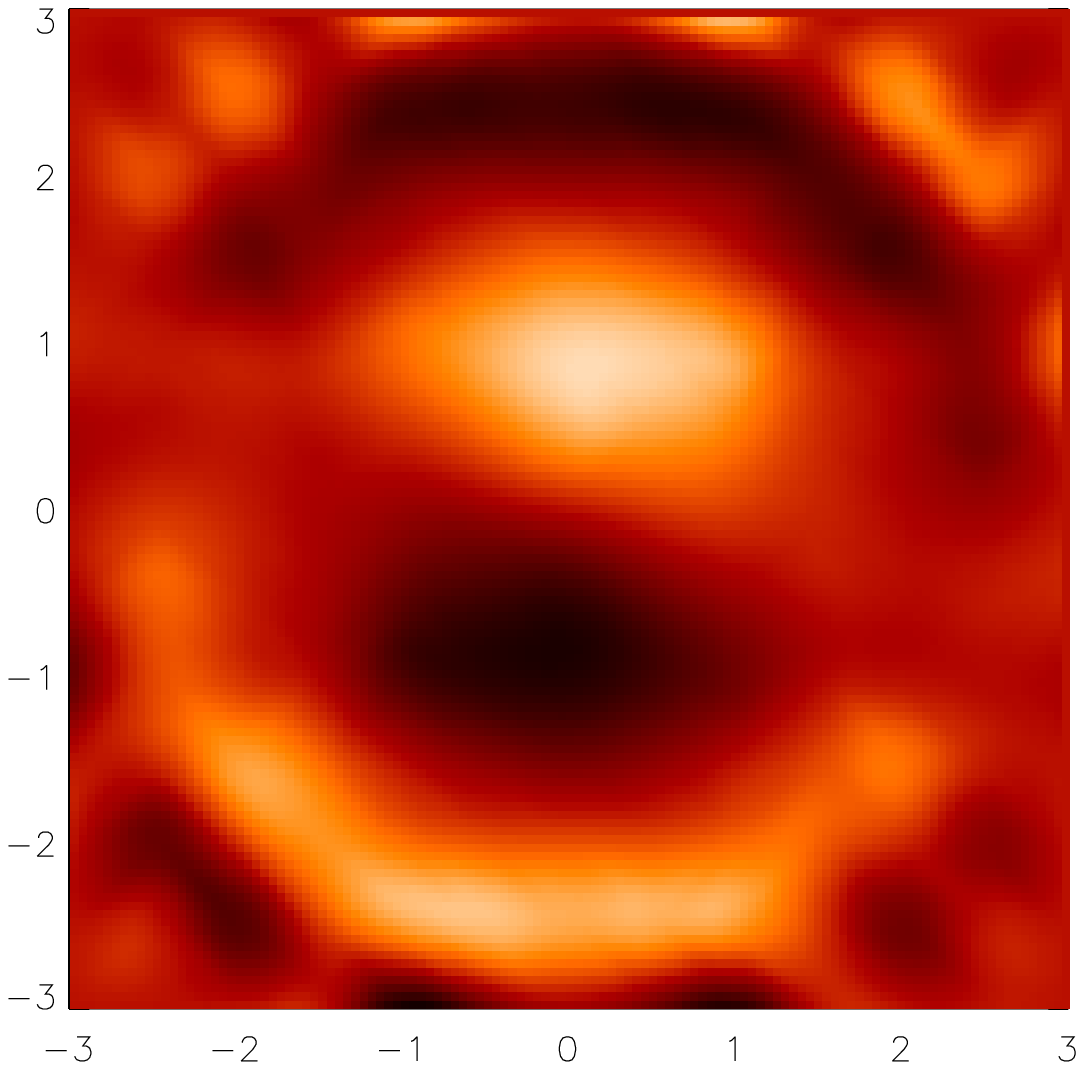}{{\it Left panel:} The mode which is most constraining in the absence of 
CMB noise, but which gets annihilated when this noise is included. This is mode 10 in Fig.~\rf{comtmp}.
{\it Right panel:} The one mode (mode 6 in Fig.~\rf{comtmp}) which
contributes significantly to the constraints when CMB noise is included.}

CMB-Cluster lensing then is sensitive not to the strength of the signal in both the hot and
the cold spots, but rather to the structure of the signal in each. This changes the calculus
of moving to lower redshift. All pixels on one side of the cluster no longer are weighted
equally, so there is no longer an advantage in the larger clusters at 
low redshift. 

\Sfig{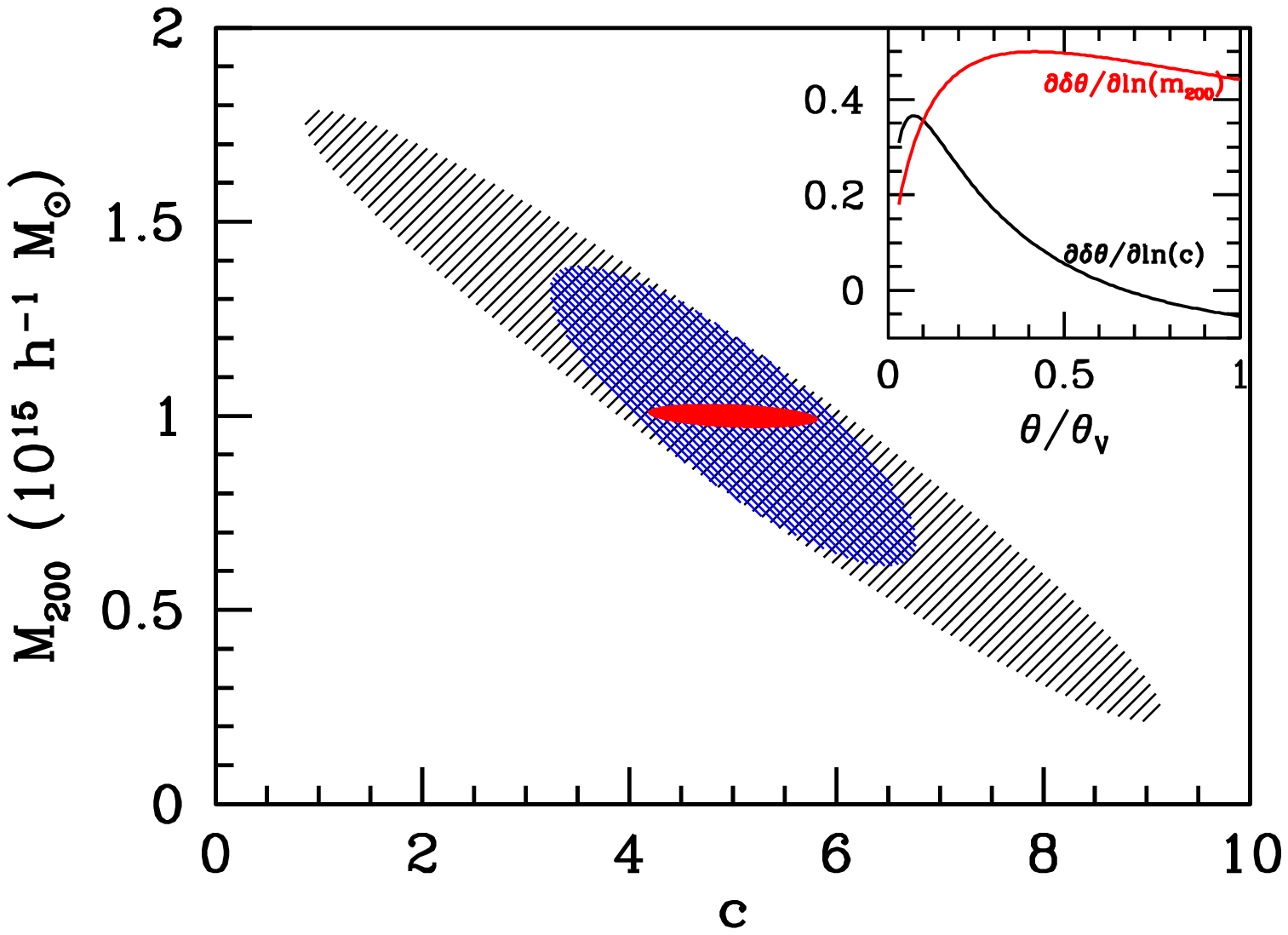}{Constraints on the mass and concentration of a $10^{15} h^{-1} M_\odot$ cluster at redshift 1 with pixel size $0.5'$ and
noise $1\mu$K. Inner solid (red) ellipse does not account for CMB noise; largest (black) ellipse does; and medium-sized
(blue) ellipse includes a $25\%
$ prior on concentration $c$. Inset in upper right demonstrates why CMB noise induces 
degeneracy between $c$ and $\mtwo$. Shown are the derivatives of the deflection angle with
respect to these two parameters. Note that the deflection angle -- and hence the signal --
is sensitive to the concentration only close to the cluster center.}

Figure~\rf{m10} shows how the constraints change on an individual cluster when CMB noise
is taken into account. The innermost ellipse shows the constraint on a massive cluster before
accounting for CMB noise. While the concentration is not pinned down with very high accuracy,
the mass is. When CMB noise is introduced, there is a degeneracy between concentration and
mass. The inset in Fig.~\rf{m10} gives a hint as to the origin of this degeneracy. The concentration
parameter affects the deflection angle, and hence the signal, only close to the cluster center.
Increasing the mass, on the other hand, increases the signal everywhere within the virial radius.
Without CMB noise, all pixels at distances $\theta\ga 0.5\theta_V$ therefore contribute to help pin
down $\mtwo$. The concentration is then determined by the innermost region, with its much smaller
number of pixels; hence the concentration is known much less accurately. In the presence of
CMB noise the situation changes: now the most important mode measures the difference between the 
signal in the inner and outer
regions. Increasing the signal in the outer region by raising the mass can be offset by increasing
the signal in the innermost region by raising the concentration. Therefore, only the sum
of the two parameters is well constrained. 

The mass-concentration degeneracy can be broken if we use the knowledge obtained from
simulations\,\cite{Jing2000,BulDek2001} about the relation between the two. Imposing a prior
on $c$ with a $25\%
$ uncertainty leads to the intermediate ellipse in Fig.~\rf{m10}. The errors on the mass of this
cluster then get reduced to useful levels. Even a more conservative prior on $c$  of $50\%
$ leaves the mass errors quite tight: the factor of two increase in the $c$ prior leads to only
a $\sim 40\%
$ increase in the mass error.

Figure~\rf{cont_cmb1d.5} shows the fractional uncertainty on cluster masses when
all forms of noise considered above are included. The errors are above
$50\%
$ for clusters with mass below $10^{15}
h^{-1} M_\odot$ even with the ambitious assumptions for noise and resolution.
Note also that the uncertainties now get larger as the cluster moves to lower redshift.
Although the cluster
abundance depends heavily on $\sigma_8$, so the numbers in Fig.~\rf{cont_cmb1d.5} (which
assume $\sigma_8=0.85$) are only a guess, it is unlikely that there will be more than a handful of
clusters for which CMB lensing will determine the mass to better than $50\%
$. One possibility for a small scale CMB experiment is to sit on these handful 
(which are first identified in optical or Sunyaev-Zel'dovich surveys). 

\Sfig{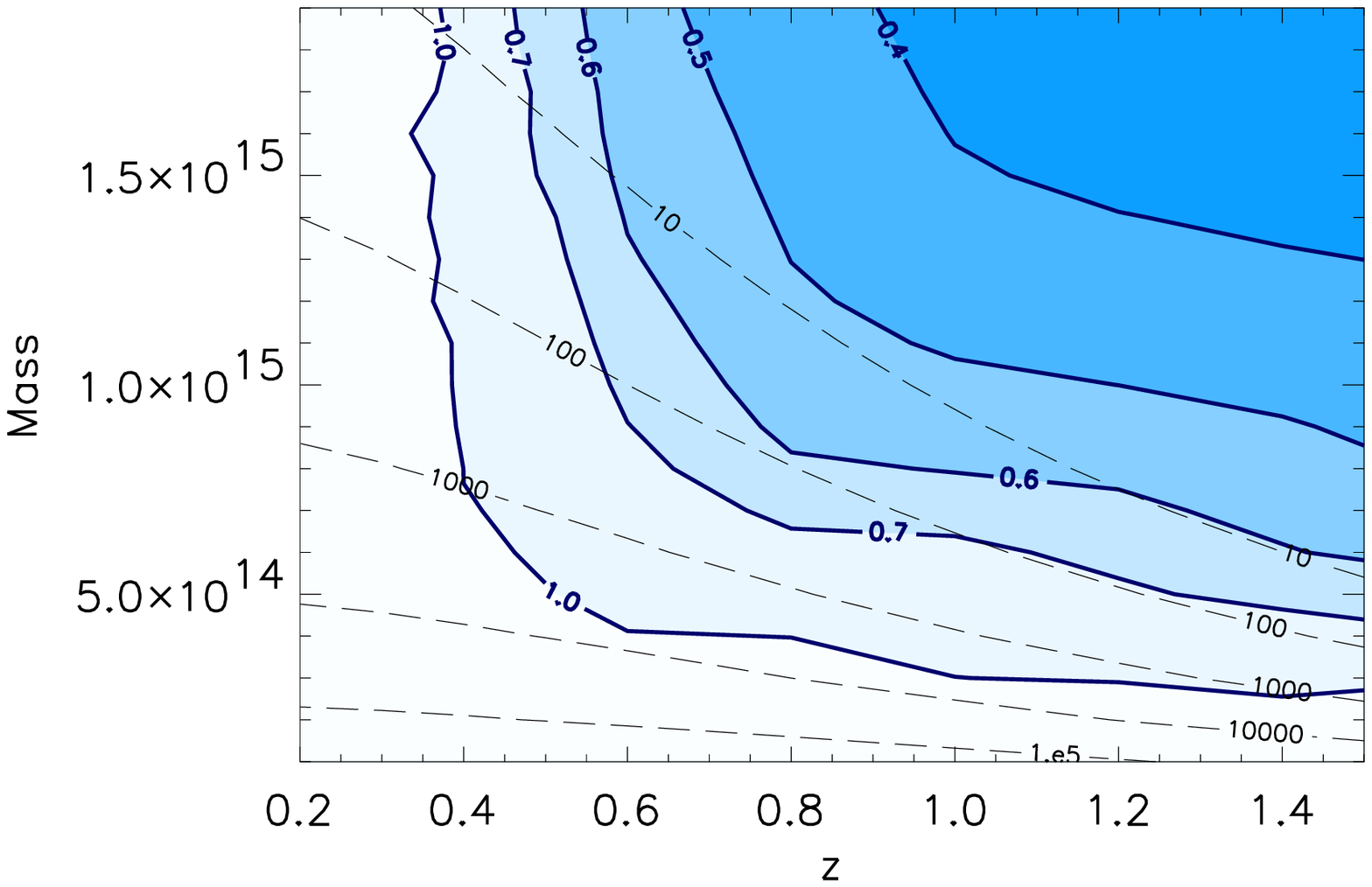}{Fractional uncertainty on cluster mass from a CMB experiment 
with pixel size $0.5'$ and noise $1\mu$K. Contours are as described in Fig.~\rf{cont_nolss}.
The constraints here include CMB noise.}

On the other hand, CMB-cluster lensing does enable us to determine a combination of the mass and concentration.
One way to exploit this is to impose a prior on the concentration. Fig.~\rf{cont_pcmbd1.5} shows the
resultant mass constraints with a $25\%
$ prior on the concentration. With this prior, we can obtain masses accurate to $50\%
$ for tens of thousands of clusters with masses above $2\times10^{14} h^{-1} M_\odot$ at redshifts above $0.4$.
The uncertainties go down to less than $30\%
$ for the hundreds of clusters at redshifts greater than $\sim 0.8$ and masses above $\sim 5\times
10^{14} h^{-1} M_\odot$. These constraints could prove very useful in determining dark energy parameters.
Alternatively, one could use the constraint on the mass/concentration in the context of a halo
model~\cite{Rozo2004}.

\Sfig{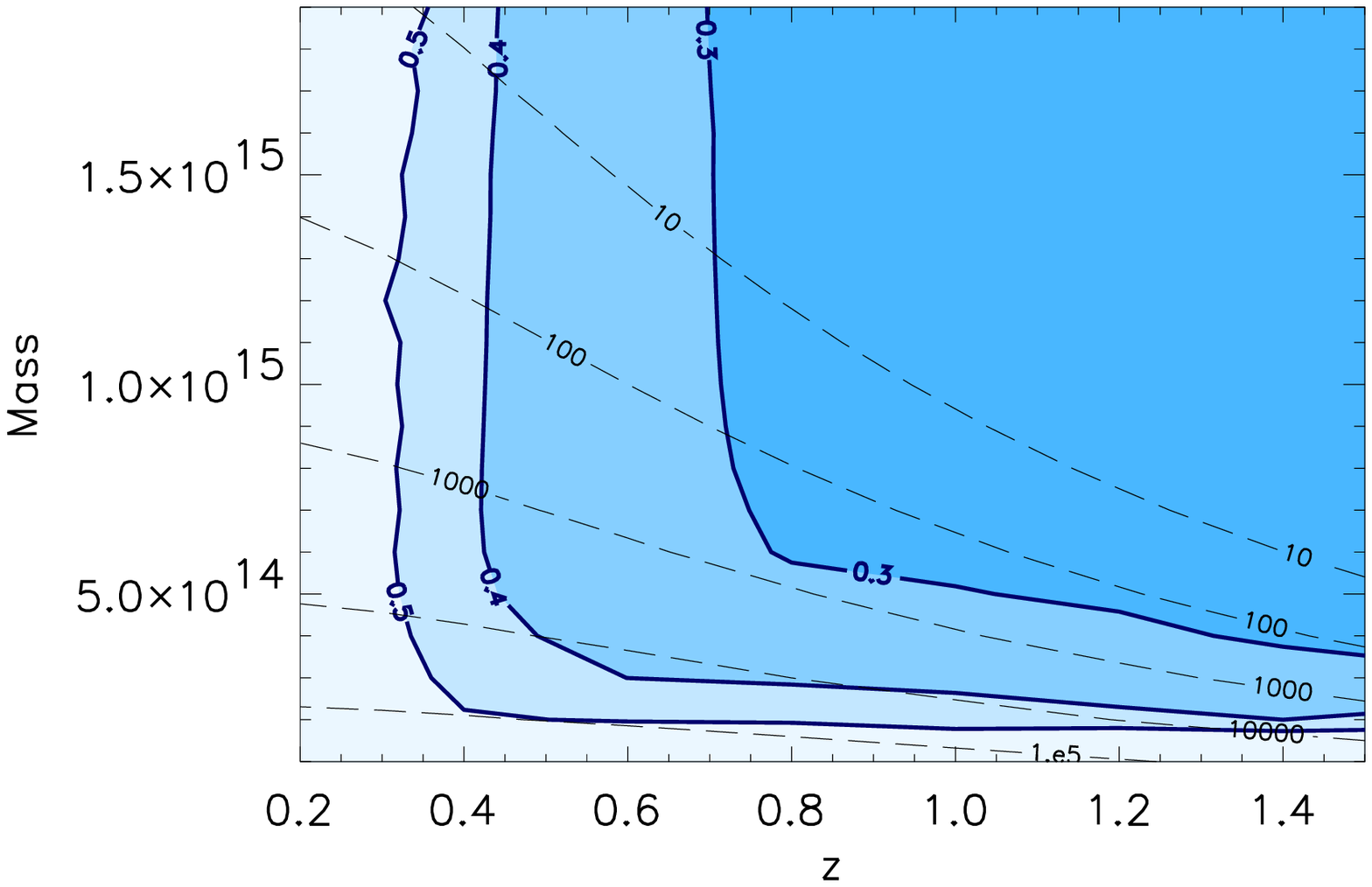}{Same as Figure \rf{cont_cmb1d.5} except that here a $25\%
$ prior is imposed on concentration $c$.}

The above constraints were for an ambitious CMB experiment with noise per pixel of $1\mu$K and $0.5'$
pixels. Figure~\rf{dncontcijp_m10z1} shows how well less ambitious experiments would do in determining
the mass of a $10^{15} h^{-1} M_\odot$ cluster at $z=1$. Apparently pixel sizes must be smaller than $\sim 1.5'$
and noise per pixel smaller than $5\mu$K for any reasonable constraints on cluster masses even including 
a concentration prior. This would seem to rule out CMB lensing in Planck as a useful tool for measuring cluster
masses.

\Sfig{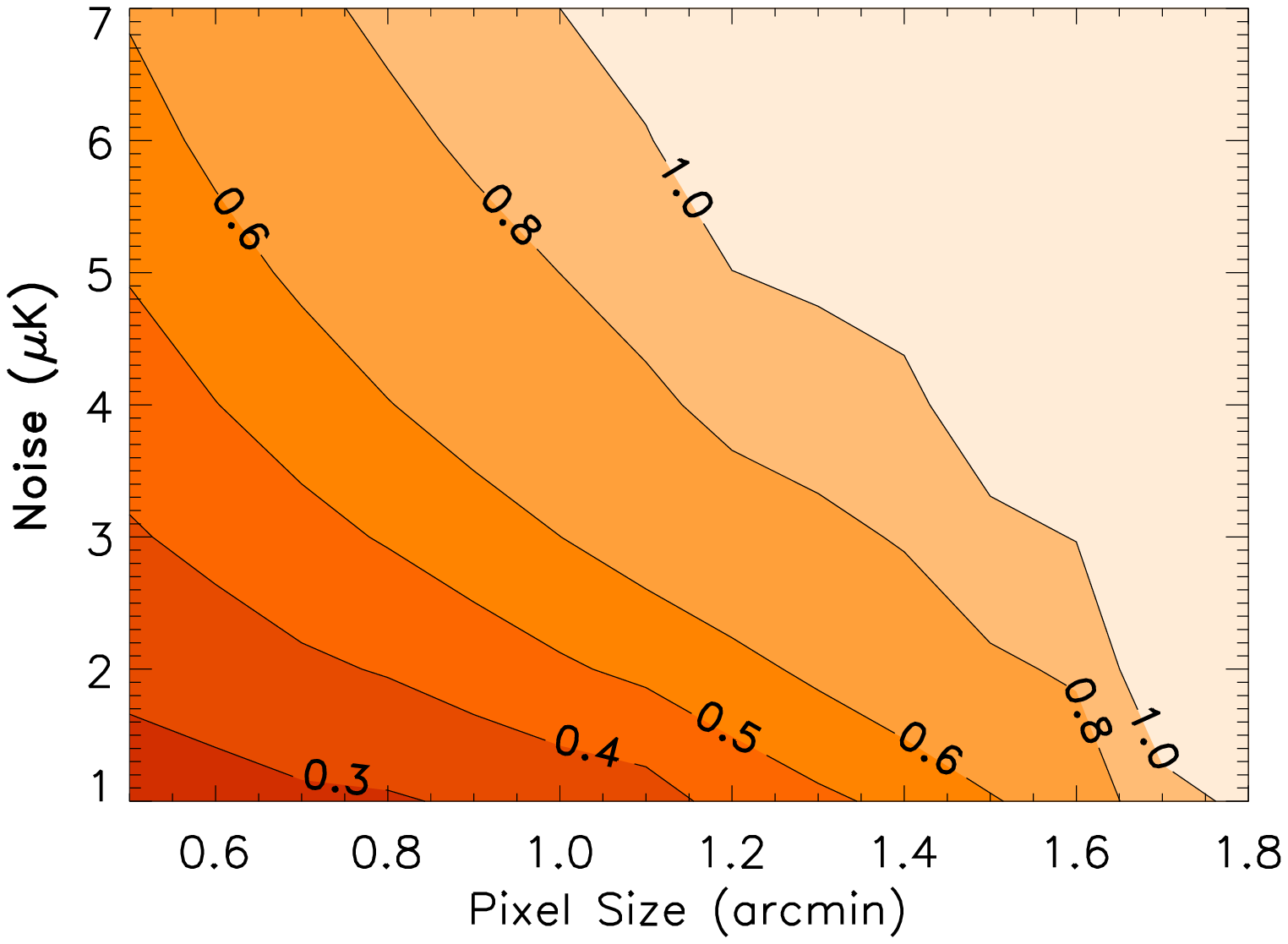}{Constraints on the mass of a cluster at redshift
1 with $\mtwo=10^{15} h^{-1}$ Mpc including CMB noise and a prior on the concentration parameter. The point
at bottom left here corresponds to the ($1,10^{15}$) point in Fig.~\rf{cont_pcmbd1.5}.}

\section{Conclusion}

Since gravitational lensing is directly sensitive to the mass of a cluster,
it seduces us into hoping that we can reject more traditional phenomenological mass indicators
such as X-ray temperature and galaxy counts. In the case considered here, lensing of the CMB,
unavoidable realities spoil some of this allure. Since the primordial CMB does have some small
scale structure, this {\it CMB noise} significantly degrades the constraining potential of
the lensed field. The constraints may still prove useful in
constraining cosmological parameters such as the dark energy equation of state, but we will have
to include this degradation when evaluating different experimental proposals.

While I believe this work is an important step on the road to assessing the utility of
CMB-cluster lensing, it is incomplete in several ways. First, I have neglected a number of other
contaminants such as point sources, the thermal Sunyaev-Zel'dovich effect, and the kinetic
Zel'dovich effect. It is at least conceivable that these other effects differ sufficiently
in frequency dependence and spatial morphology from the cluster lensing signal that
they will not further degrade the projected constraints much. For example, 
if the morphology of the kinetic Sunyaev-Zel'dovich effect is much different from
that depicted in the right panel of Fig.~\rf{u17}, then it would have no impact on
the mass determination, for we have seen that this is the mode which most tightly
constraints the mass. Second, I have assumed a particularly
simple parametrized form for the cluster mass distribution. Accounting for asphericity and substructure
requires numerical simulations~\cite{HolKos,Vale2004}, and these effects might add further noise to the
mass estimators. Note that all of these effects would serve only to degrade the mass constraints further.

To close on a positive note, I point out that CMB experiments have consistently exceeded our expectations.
There is thus every reason to believe that the high resolution, low noise experiments anticipated here
will take place. If they do, then -- along with input from numerical simulations such as the correlation
between mass and concentration -- we can hope to obtain interesting limits on the masses of thousands of
clusters from CMB maps.

{\it Note:} After this work was completed (but before it was circulated), 
Holder and Kosowsky~\cite{HolKos} and Vale, Amblard, and White~\cite{Vale2004}
put out preprints studying very similar issues.
Although we reached the same general conclusion -- realistic effects degrade
the ability of CMB-cluster lensing to determine masses -- 
our papers are by and large complementary.
While I have used analytic techniques and therefore simplified cluster models, the
other two papers use more realistic numerical simulations. The simplicity has allowed
me to provide quantitative measures of the efficiency of CMB-cluster lensing
for a wide range of cluster masses, redshifts, noise levels, and pixel sizes.
Ultimately, these estimates will need to be reinforced with the realistic simulations
performed in \cite{HolKos,Vale2004}.

\begin{acknowledgments}
I am very grateful to Eduardo Rozo for very helpful suggestions and
to Gil Holder and Arthur Kosowsky for useful conversations.
This work is supported by the DOE, by NASA grant NAG5-10842, 
and by NSF Grant PHY-0079251. 
\end{acknowledgments}

\bibliography{v2}
\end{document}